\documentclass[reqno]{amsart}
\usepackage{fullpage,graphicx,psfrag,epsfig,amsmath,amsfonts,amssymb,fancybox,amsthm}
\usepackage[dvips]{color}

\newcommand{\eq}{\begin{equation}}
\newcommand{\en}{\end{equation}}
\newcommand{\eqa}{\begin{eqnarray}}
\newcommand{\ena}{\end{eqnarray}}
\newcommand{\eqas}{\begin{eqnarray*}}
\newcommand{\enas}{\end{eqnarray*}}
\newcommand{\ra}{ {\rightarrow} }

\newcommand{\bt}{\beta}

\newcommand{\p}{\partial}
\newcommand{\ep}{\epsilon}
\newcommand{\no}{\nonumber}
\newcommand{\mc}{\mathcal}
\newcommand{\mb}{\mathbb}

\newcommand{\mbf}{\mathbf}

\newcommand{\f}{\frac}
\newcommand{\stm}{\setminus}

\newtheorem{theorem}{Theorem}[section]
\newtheorem{lemma}[theorem]{Lemma}
\newtheorem{claim}[theorem]{Claim}

\theoremstyle{definition}

\theoremstyle{remark}
\newtheorem{remark}[theorem]{Remark}

\numberwithin{equation}{section} \numberwithin{figure}{section}
\def\mb#1{\mathbf{#1}}

\begin{document}
\title{A rigorous proof of the cavity method for counting matchings \footnote{This work was supported by Microsoft research.}
}

\author{Mohsen Bayati$^1$ and Chandra Nair$^2$}
\address{$^1$Department of Electrical
Engineering, Stanford University,
        350 Serra Mall, Stanford, CA 94305, USA
        {\tt\small bayati@stanford.edu}}%
        \address{$^2$Theory group, Microsoft Research,
        One Microsoft Way, Redmond, WA 98052
        {\tt\small cnair@microsoft.com}}%

\maketitle \thispagestyle{empty} \pagestyle{empty}

\begin{abstract}
In this paper we rigorously prove the validity of the cavity
method for the problem of counting the number of matchings in
graphs with large girth. Cavity method is an important heuristic
developed by statistical physicists that has lead to the
development of faster distributed algorithms for problems in
various combinatorial optimization problems. The validity of the
approach has been supported mostly by numerical simulations. In
this paper we prove the validity of cavity method for the problem
of counting matchings using rigorous techniques. We hope that
these rigorous approaches will finally help us establish the
validity of the cavity method in general.
\end{abstract}

\section{INTRODUCTION}

\subsection{Motivation}

Distributed message passing algorithms like belief propagation
have been around for over a decade now \cite{Pearl1988,
YFW00-22,Wainwright2003,McEliece2000}. Recently some important
problems in combinatorial optimization have seen faster
distributed algorithms - motivated using a heuristic technique in
statistical physics called the cavity method - that seem to solve
problem instances much larger than what was previously feasible.
This method has also led to analytical predictions about some
threshold phenomenon in various problems of cross-disciplinary
interest. Some examples of its application include the
satisfiability threshold for random constraint satisfaction
\cite{MezardParisiZecchina,MezardZecchina} and the corresponding
survey-propagation algorithm , iterative decoding algorithms in
multi-user CDMA \cite{kab2005}, etc.

However, very few rigorous results are known concerning the
validity of the cavity method and the convergence of the
algorithms. In this paper we wish to add to the body of rigorous
results \cite{Aldous,Bayati,GamarnikNowicki,Talagrand,Urbanke}
supporting the predictions of the cavity method by showing its
correctness for the problem of counting matchings in large sparse
graphs. We borrow some of the techniques from Gamarnik et.al.
\cite{GamarnikNowicki} but we hope that some newer lines of the
argument (e.g. showing validity of the free energy shifts) could
lead to useful insights into the validity of the method for other
instances in which the method was applied.

The algorithms generated using this method in some instances
resemble the naive believe propagation equations, whereas in some
other instances they resemble two-layered belief propagation (or
survey propagation) equations, and in few other cases are
significantly more involved. In the problem of counting matchings,
the equations generated using the cavity method resemble the naive
belief propagation equations. In this paper we show the
convergence of the cavity equations and the uniqueness of the
fixed points for arbitrary graphs $G$. In general such convergence
results are not known except for trees or graphs with exactly one
cycle \cite{Weiss1997}.

\subsection{The matching problem}

Counting the number and size of matchings on various types of
random graphs has been a classical problem in graph theory. This
problem has been intensively studied for a long time by
mathematicians and computer scientists \cite{MicaliVazirani}. Very
recently Zdebrov\'{a} and M\'{e}zard \cite{ZdeborovaMezard} used
the cavity method to solve this problem. They believed that the
results obtained by this heuristic are exact for the matching
problem and using this approach: a) they derived an algorithm that
computes the entropy for arbitrary graphs with girth that diverges
in the large size limit and b) derived analytical results for
regular and Erd\"{o}s-R\'{e}nyi random graph ensembles.

We first define the problem of finding the number of perfect
matchings in a simple graph and then describe the cavity equations
for solving it.

\subsubsection{Problem setup and notation}

Consider a graph $G=(V,E)$ with $n$ vertices $V$, and edge-set
$E$. Throughout this paper we will always assume that $G$ is
simple (i.e. G has no multi-edge or self-loops) and undirected.
The {\em girth} of a graph $G$ is defined as the length of the
shortest cycle. A {\em matching} is a subset of edges $M\subset E$
such that no two edges of $M$ have a common endpoint.  Let $|M|$
denote the size of matching $M$ and let $M^*$ be a matching of
maximum size. If $|M^*|=n/2$ then $M^*$ is called a {\em perfect}
matching.

Counting the number of perfect matchings in a graph $G$ is shown
to be \#-P complete \cite{Valiant}, (i.e. in general no
polynomial-time algorithm can find the exact number of perfect
matchings of $G$ unless $P=NP$). Since it is widely believed that
$P \neq NP$, many approaches have been focused on finding
polynomial-time algorithms for approximately counting the number
of perfect matchings
\cite{JerrumSinclair,JerrumSinclairVigoda,FurerKasiviswanathan,Chien}.

Let $a,b,\ldots$ denote the vertices of $G$  and $i,j,\ldots$ denote
the edges. For every vertex $a \in G$ let $N(a)$ denote the
vertex-neighborhood of vertex $a$, i.e. $N(a) = \{ b: (a,b) \in E
\}$, and let $E(a)$ denote the edge-neighborhood of vertex $a$, i.e.
the set of edges in $E$ that have the vertex $a$ as an endpoint.

Describe a matching $M$ by variables $s_i=s_{a,b}\in\{0,1\}$
assigned to each edge $i=(a,b)\in E$ with
$$
s_i=\left\{
\begin{array}{ll}
1&\textrm{edge }i\in M\\
0&\textrm{edge }i\notin M
\end{array}
\right.
$$

Since $M$ is a matching, it follows that for any vertex $a\in V$:
\[ \sum_{b \in  N(a)}s_{a,b}\leq 1 \]

For every matching $M \subset E$ define its energy to be its
number of unmatched vertices:
 \[E_G(M)=\sum_{a\in V}E_a(M)=n-2|M|\]
where $E_a(M)=1-\sum_{b \in N(a)}s_{a,b}$.

This induces a probability distribution (called the {\em Gibb's}
distribution) on the set of all matchings, $\mc{M}(G)$,  of the
graph $G$, defined by:
\[P_{G,\bt}(M)=\f{1}{Z_G(\bt)}e^{-\bt E_G(M)}\]
where $\bt$ is a positive number and is called  {\em inverse
temperature}. The normalizing term $Z_G(\bt)=\sum_{M}e^{-\bt
E_G(M)}$ is called {\em partition function}.

The partition function of an empty graph is defined as $1$. To
simplify notation we will omit the dependence on $\bt$ and write
$P_G$ instead of $P_{G,\bt}$ and $Z_G$ instead of $Z_G(\bt)$.

\medskip

Let a configuration denote a collection of edges $S \subset E$.
Observe that $P_G$ can also be represented on the set of all
configurations, as below. Let for all $a\in V$
\[
\psi_a(S)=\mbf{1}_{\{\sum_{b \in N(a)}s_{a,b}\leq
1\}}e^{-\bt(1-\sum_{b \in N(a)}s_{a,b})}.
\]
Then it follows that
\[
P_G(S)=\f{1}{Z_G}\prod_{a\in V}\psi_a(S).
\]
Note that $E_G(M)$ is minimized when $M$ is a maximum size
matching and for large values of $\bt$ the partition function is
dominated by the terms corresponding to maximum size matchings of
$G$. Hence for $\bt\gg 0$

\[ F_G \stackrel{\triangle}{=} -\f{1}{\bt}\log Z_G \approx \f{-\log(N_G)}{\bt}+E_G(M^*)\]
where $N_G$ is number of maximum size matchings in $G$. $F_G(\bt)$
is defined as the free-energy of the system. Observe that for large
$\bt$, we have
\begin{eqnarray*}
E_G(M^*) &\approx &\frac{\partial}{\partial \bt}(\bt F_G)\\
\log(N_G) &\approx & \beta\left(\frac{\partial}{\partial \bt}(\bt
F_G) - F_G\right),
\end{eqnarray*}
with the approximation becoming exact as $\beta \to \infty$.

The cavity method of statistical physics is a heuristic that is used
to evaluate the partition function. In the next section we will
state the equations derived in \cite{ZdeborovaMezard} using the
cavity approach and then prove that these equations compute the
partition function exactly for sparse graphs.

\begin{remark}
In the first few sections the term sparse graph is used loosely to
mean graphs that have no short loops. The precise dependence
needed between the length of the shortest loop and the size of the
graphs (measured in terms of the number of vertices and number of
edges) will be spelt out in the final section.
\end{remark}

\subsection{The cavity-claims for the matching problem}

Let $h^{i \to a} : E \times V \to \mathbb{R}$ be the `message'
that edge $i= (a,b)$ conveys to vertex $a$, one of its end-points.
Note that there are $2|E|$ messages as there are two messages for
each edge.

The following two claims form the algorithmic and analytical crux
of the cavity method for this problem.

\begin{claim}
[Zdeborov\'{a}-M\'{e}zard] \label{cl:bp}
Consider the iterative equation defined by
\begin{equation}
\label{eq:iter}
h^{i \to a}(t+1) = - \frac{1}{\bt} \log \left[ e^{-\bt} + \sum_{j \in
E(b)\setminus i} e^{\bt h^{j \to \bt}(t)} \right].
\end{equation}

These iterative equations converge to a unique fixed point for a
large sparse graph whose girth diverges with the size of the
graph.
\end{claim}

Let $h^{i \to a}$ be the unique fixed points of the system of
equations in the above claim, i.e.
\begin{equation}
\label{eq:recur} h^{i \to a} = - \frac{1}{\bt} \log \left[
e^{-\bt} + \sum_{j \in E(b)\setminus i} e^{\bt h^{j \to \bt}}
\right].
\end{equation}

\begin{claim}
[Zdeborov\'{a}-M\'{e}zard] \label{cl:cavity} The free energy for a
single large sparse graph is given by
\[ F_G = \sum_a \Delta F_a - \sum_i \Delta F_i \]
where
\begin{equation*}
e^{-\bt \Delta F_a} =  e^{-\bt} + \sum_{i \in E(a)} e^{\bt h^{i
\to a}} , \quad e^{-\bt \Delta F_i}  =  1 + e^{\bt (h^{i \to a} +
h^{i \to b})}.
\end{equation*}
\end{claim}

$\Delta F_a$ is often called the {\em free energy shift}
corresponding to the removal of a vertex $a$ and its associated
edges, and $\Delta F_i$ the {\em free energy shift} corresponding
to the removal of the edge $i$.
\medskip

\begin{remark}
Our outline of the proof is as follows: first we will prove the
cavity-claims for the case when $G$ is a tree, and then we will
proceed to establish this for a large sparse graph. Validity of
Claim \ref{cl:bp} is well-known for the case of a tree and the
main result in the next section is to show that Claim
\ref{cl:cavity} is exact as well.
\end{remark}

\section{The validity of the cavity-claims on a tree}

In this section, we shall assume that the graph $G$ is a tree.
Note that removing  any edge $i=(a,b)$ from $G$ splits the graph
into two subgraphs: $G^{i,a}$ containing the vertex $a$, and
$G^{i,b}$ containing the vertex $b$.

\begin{figure*}
\centering
      \includegraphics[scale=0.5]{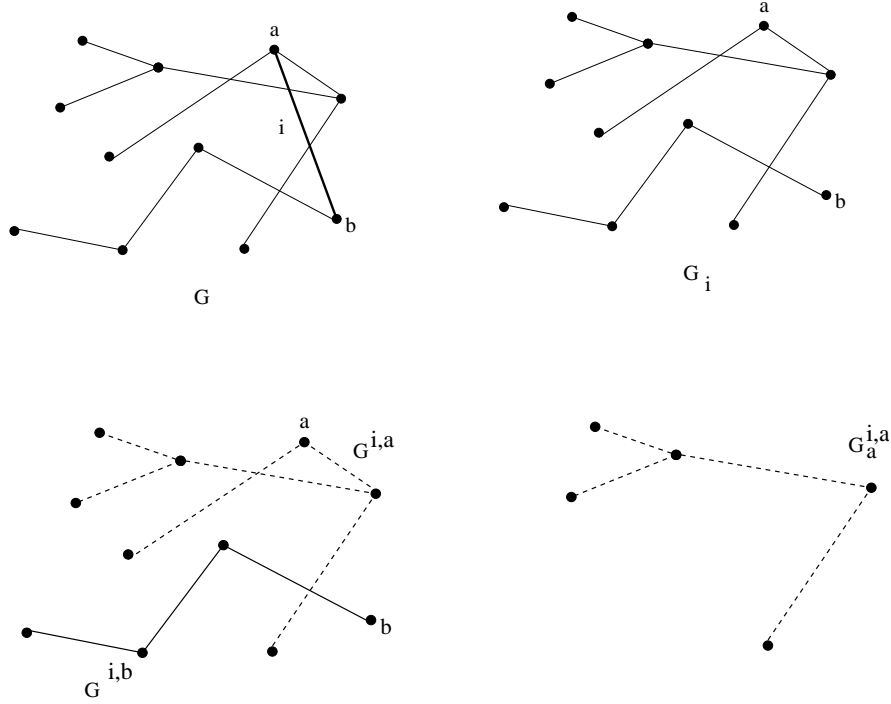}
      \caption{The graph $G$ and the various sub-graphs}
      \label{fig:graph}
\end{figure*}

\begin{remark}
For any graph $G=(V,E)$ with $a\in V$ and $i\in E$, let $G_i$ denote
the graph with edge $i$ removed. Further, let $G_a$ denote the graph $G$
with vertex $a$ and  with all edges adjacent to $a$ removed. In Figure
\ref{fig:graph}, observe that $G^{i,a}_a$ is formed from $G^{i,a}$ by removing
vertex $a$ and all its adjacent edges.
\end{remark}

Observe that,
\begin{equation*}
\f{Z_{G_i}}{Z_G} = \f{Z_{G^{i,a}} Z_{G^{i,b}}}{Z_{G^{i,a}} Z_{G^{i,b}} + Z_{G_a^{i,a}} Z_{G_b^{i,b}}}
\end{equation*}

Defining
\begin{equation}
e^{\bt h^{i \to b}} =  \f{Z_{G_a^{i,a}}}{Z_{G^{i,a}}},
\qquad e^{\bt h^{i \to a}} = \f{Z_{G_b^{i,b}}}{Z_{G^{i,b}}}, \label{eq:var}
\end{equation}
we see that
\begin{equation}
\label{eq:feshftedge} e^{-\bt \Delta F_i} \stackrel{\triangle}{=}
\f{Z_G}{Z_{G_i}} = 1 + e^{\bt (h^{i \to a} + h^{i \to b})}.
\end{equation}
Note that
\begin{equation*}
\f{Z_{G_a}}{Z_G} = \f{\prod_{i \in E(a)} Z_{G^{i,b_i}}}{\prod_{i
\in E(a)} Z_{G^{i,b_i}}( e^-\bt + \sum_{i \in E(a)}
\frac{Z_{G_{b_i}^{i,b_i}}}{Z_{G^{i,b_i}}})}.
\end{equation*}
Thus,
\begin{equation}
\label{eq:feshftvrtx} e^{-\bt \Delta F_a} \stackrel{\triangle}{=}
\f{Z_G}{Z_{G_a}} = e^{-\bt} + \sum_{i \in E(a)} e^{\bt h^{i \to
a}}.
\end{equation}

\begin{lemma}
\label{le:fshfttree} Free energy $F_G$, can be expressed as the
sum of free energy shifts when G is a tree, i.e.
\[ F_G = \sum_a \Delta F_a - \sum_i \Delta F_i \]
\end{lemma}
\begin{proof}
Observe that
\begin{equation}
\begin{aligned}
& \prod_{a \in V} e^{-\bt \Delta F_a}\prod_{i\in E} e^{\bt \Delta
F_i} = \prod_{a\in V} \f{Z_G}{Z_{G_a}} \prod_{i \in E}
\f{Z_{G_i}}{Z_G} \\
& \quad = \prod_{a\in V} \f{Z_G}{\prod_{i \in E(a)} Z_{G^{i,b_i}}}
\prod_{i \in E} \f{Z_{G^{i,a}} Z_{G^{i,b}}}{Z_G}
 \stackrel{(a)}{=} \f{Z_G^{|V|}}{Z_G^{|E|}} \quad
\stackrel{(b)}{=} Z_G = e^{-\bt F_G}
\end{aligned}
\end{equation}
Here $(a)$ follows form the fact that
\[ \prod_{a \in V} \prod_{i \in E(a)} Z_{G^{i,b_i}} = \prod_{i \in E} Z_{G^{i,a}} Z_{G^{i,b}} \]
and $(b)$ follows form the fact that in a tree $|E| = |V| - 1$.
\end{proof}

To complete the proof of Claim \ref{cl:cavity} for the case that G is a tree we need to show the following:
\begin{itemize}
\item[$(i)$] The variables $h^{i \to a}$ defined in \eqref{eq:var} satisfy equation \eqref{eq:recur}.
\item[$(ii)$] The equations \eqref{eq:recur} have a unique fixed
point.
\end{itemize}

\begin{remark}
The fact that the equations \eqref{eq:recur} have a unique fixed
point is a well-known fact for the case when $G$ is a tree. For
general graphs $G$, the convergence and the uniqueness is not
known.  One of the main technical ingredients in this paper is to
establish  the convergence and the uniqueness for general graphs
as well.
\end{remark}

\begin{lemma}
\label{le:recsatisfy} The variables $h^{i \to a}$ defined in \eqref{eq:var}
satisfy equation \eqref{eq:recur}.
\end{lemma}

\begin{proof}
 We need to show that
 \[
 h^{i \to a} = - \frac{1}{\bt} \log \left[ e^{-\bt} + \sum_{j \in
E(b)\setminus i} e^{\bt h^{j \to \bt}} \right]. \]
This is equivalent to showing that
\[ e^{-\bt h^{i \to a}} + e^{\bt h^{i \to b}} = e^{-\bt} + \sum_{j \in E(b)}
e^{\bt h^{j \to b}}.
\]
Now using the equations
\eqref{eq:feshftvrtx}, \eqref{eq:var} we see that this reduces to
showing
\[
\f{Z_{G^{i,b}}}{Z_{G_b^{i,b}}} + \f{Z_{G_a^{i,a}}}{Z_{G^{i,a}}} =
\f{Z_G}{Z_{G_b}}.
\]
Observe that this follows from the following: $Z_G =
Z_{G^{i,b}}Z_{G^{i,a}}+ Z_{G_b^{i,b}}Z_{G_a^{i,a}}$ and  $Z_{G_b}
= Z_{G^{i,a}}Z_{G_b^{i,b}}$.

\end{proof}

\section{Convergence of the iterative equations}
\label{sec:conv}

Consider any simple graph $G$ and consider the iterative equations
defined on it according to \eqref{eq:iter}. The proof for the
convergence of the iterative equations is based on the following
lemma:

\begin{lemma}\label{lem:Lip}
Let $f:\mb{R}^{rs}\longrightarrow \mb{R}$ be a real valued function
defined as follows:
\begin{equation}
\label{eq:function1}
f(\textbf{x})
=\f{-1}{\bt}\log{\Big(e^{-\bt}+\sum_{k=1}^r\f{1}{e^{-\bt}+\sum_{\ell=1}^s
e^{\bt x_{k\ell}}}\Big)}
\end{equation}
then for any $\textbf{x}, \textbf{y}\in\mb{R}^{rs}$:
\begin{equation}
|f(\textbf{x})-f(\textbf{y})| \leq \f{r}{r+e^{-2\bt}}
\|\textbf{x}-\textbf{y}\|_\infty. \label{eq:Lip}
\end{equation}
\end{lemma}

\begin{proof}
Since $f(\textbf{x})$ is differentiable,  multi-variable version
of the mean value theorem implies that for any $\textbf{x},
\textbf{y}\in\mb{R}^{rs}$ there exist a point $\textbf{z}$ on the
line-segment connecting $\textbf{x}$ and $\textbf{y}$ in
$\mb{R}^{rs}$ such that:
\begin{equation*}
f(\textbf{x})-f(\textbf{y})=\nabla
f(\textbf{z})\cdot(\textbf{x}-\textbf{y})
\end{equation*}
From H\"{o}lder's inequality it follows that
\[
|f(\textbf{x})-f(\textbf{y})| \leq \|\nabla f(\textbf{z})\|_1
\|\textbf{x}-\textbf{y}\|_\infty
\]

\medskip

In order to show (\ref{eq:Lip}) it suffices to show that $\|\nabla
f(\textbf{z})\|_1 \leq \f{r}{r+e^{-2\bt}}$. Note that,
\begin{equation}
\|\nabla f(\textbf{z})\|_1=\sum_{k,\ell}|\f{\p f}{\p z_{k\ell}}|
=\sum_{k,\ell}\f{\f{ e^{\bt
z_{k\ell}}}{(e^{-\bt}+\sum_{q=1}^se^{\bt
z_{kq}})^2}}{e^{-\bt}+\sum_{p=1}^r\f{1}{e^{-\bt}+\sum_{q=1}^s
e^{\bt z_{pq}}}} =\sum_{k=1}^r\f{\f{\sum_{\ell=1}^se^{\bt
z_{k\ell}}}{(e^{-\bt}+\sum_{q=1}^se^{\bt
z_{kq}})^2}}{e^{-\bt}+\sum_{p=1}^r\f{1}{e^{-\bt}+\sum_{q=1}^s
e^{\bt z_{pq}}}}.\no
\end{equation}
For simplicity of notation let $A_k =\f{1}{e^{-\bt}+
\sum_{\ell=1}^se^{\bt} z_{k\ell}}$, then one obtains
\begin{equation*}
\|\nabla f(\textbf{z})\|_1 =
\f{\sum_{k=1}^r(1-e^{-\bt}A_k)A_k}{e^{-\bt}+\sum_{k=1}^rA_k}  =
1-\f{e^{-\bt}+e^{-\bt}\sum_{k=1}^rA_k^2}{e^{-\bt}+\sum_{k=1}^rA_k}.
\end{equation*}
Now using $0\leq A_k\leq e^\bt$:
\begin{eqnarray}
\|\nabla f(\textbf{z})\|_1&\leq&
1-\f{e^{-\bt}}{e^{-\bt}+re^{\bt}}=\f{r}{r+e^{-2\bt}}\no
\end{eqnarray}
This completes the proof of Lemma \ref{lem:Lip}.
\end{proof}

Next we will use Lemma \ref{lem:Lip} to prove convergence of the cavity
equations (\ref{eq:iter}) for any graph $G$.

\begin{theorem}
\label{th:maincon}
For any graph $G$ the set of cavity equations (\ref{eq:iter}) converges to a
unique fixed point independent of its initial conditions.
\end{theorem}
\begin{proof}
Consider an arbitrary initial condition $\{h^{i\ra a}(0)\}$.
The iterative equations in \eqref{eq:iter} states that
\[ h^{i \to a}(t+1) = - \frac{1}{\bt} \log \left[ e^{-\bt} + \sum_{j \in
E(b)\setminus i} e^{\bt h^{j \to \bt}(t)} \right]. \]

Define $F(\textbf{x})$ to be the multi-valued function from $\mb{R}^{2|E|}$ to
$\mb{R}^{2|E|}$ such that
\[ \{h^{i\ra a}(t+1)\}=F(\{h^{i\ra a}(t)\}).\]

Consider the two-iterate of function $F$, i.e. let $F^2=F\circ F$. Let
$F^2=(f_1(\textbf{x}),\ldots,f_{2|E|}(\textbf{x}))$ where each $f_i$
is a real valued function on $\mb{R}^{2|E|}$.

Observe that each function $f_i(\textbf{x})$ can be written in the form
$\eqref{eq:function1}$ where $r,s\leq \max_{a\in V}($deg$(a))$.
Let $\Delta = \max_{a\in V}($deg$(a))$, the maximum degree of a vertex in $G$.
Now using Lemma \ref{lem:Lip} for any $t\geq 2$, we have:

\begin{equation}
 \begin{aligned}
& \|\{h^{i\ra a}(t+2)\}-\{h^{i\ra a}(t)\}\|_\infty
= \|F^2(\{h^{i\ra a}\}(t))-F^2(\{h^{i\ra a}(t-2)\})\|_\infty\\
&\quad =\max_{1\leq k\leq 2|E|}(|f_k(\{h^{i\ra
a}(t)\})-f_k(\{g^{i\ra a}(t-2)\}|)
 \leq\f{\Delta}{e^{-2\bt}+\Delta}\|\{h^{i\ra a}(t)\}-\{h^{i\ra
a}(t-2)\}\|_\infty\\
&\quad \leq \left(\f{\Delta}{e^{-2\bt}+\Delta}\right)^{t/2}\|\{h^{i\ra
a}(2)\}-\{h^{i\ra
a}(0)\}\|_\infty.
 \end{aligned}
\label{eq:attract}
\end{equation}

Thus the sequence $\{h^{i\ra a}(t)\}$ is Cauchy and
hence converges to a point $\{h^{i\ra a}\}\in \mb{R}^{2|E|}$. This shows that
the equations \eqref{eq:iter} converge for any graph $G$.

To show uniqueness consider two different initial conditions $\{h^{i\ra
a}(0)\}$ and $\{g^{i\ra a}(0)\}$.
Using the same argument as in equation (\ref{eq:attract}) one has:
\begin{equation}
 \begin{aligned}
& \|\{h^{i\ra a}(t)\}-\{g^{i\ra a}(t)\}\|_\infty \\
&\quad \leq ~~~ \left(\f{\Delta}{e^{-2\bt}+\Delta}\right)^{t/2}\|\{h^{i\ra
a}(0)\}-\{g^{i\ra a}(0)\}\|_\infty
 \end{aligned}
 \label{eq:diffbc}
\end{equation}
so both sequences $\{h^{i\ra a}(t)\}$ and $\{g^{i\ra a}(t)\}$
converge to the same point and this contradicts the existence of
multiple fixed points.

This completes the proof of the Theorem \ref{th:maincon} and shows
that the equations \eqref{eq:iter} converge to a unique fixed
point for any graph.
\end{proof}

In the next section we show that validity of the cavity-claims
when the graph $G$ has a large girth.

\section{ Validity of the Cavity-Claims for graphs with large girth }

Theorem \ref{th:maincon} proves the validity of the Claim
\ref{cl:bp} for arbitrary graphs and in particular for graphs with
large girth. Therefore it suffices to show the validity of Claim
\ref{cl:cavity} for graphs with large girth. Before starting the
proof, we note the following bounds on the values of the fixed
points of the equation \eqref{eq:iter}. The proof of this lemma is
straightforward and is omitted.

\begin{lemma}
\label{le:bound} Let $\{h^{i \to a}\}$ be the the unique fixed
points of the iterative equations on any graph $G$ with maximum
degree $\Delta$. Then
\[ -\f{1}{\bt} \log \left[ e^{-\bt} + (\Delta - 1) e^\beta \right] \leq h^{i
\to a} \leq 1. \]
\end{lemma}

Consider a fixed graph $G$ of size $n$. Let $r_a$ denote the
maximum distance such that the subgraph, $G(a;r_a)$,  formed using
the vertices that are within a  distance $r_a$ from the vertex $a$
is a tree. Let the vertices in $G(a;r_a)$ be denoted by
$V(a;r_a)$. Consider the set of edges, $C$, that connect
$V(a;r_a)$ and $V(a;r_a)^c$. Further let the subgraph formed by
the vertices $V(a;r_a)^c$ be denoted as $H(a;r_a)$.

This decomposes the original graph $G$ into three parts: the
subgraph $G(a;r_a)$, the set of edges $C$, and the subgraph
$H(a;r_a)$. Pick any subset of the edges $D \subset C$. Let
$\mc{M}_{D} \subset \mc{M}(G)$ denote the set of matchings in $G$
that use precisely the subset of edges $D$ in $C$. Let $V_C
\subset V(a;r_a)$ denote the set of vertices in $G(a;r_a)$ that
are the endpoints of the edges $C$. Let $V_D \subset V_C$ denote
the set of vertices in $G(a;r_a)$ that are the endpoints of the
edges $D$. Further, let $U_D$ denote the set of vertices in
$H(a;r_a)$ that are the endpoints of the edges $D$.

Denote $G(a;r_a - 1)$ as the sub-graph formed using the vertices
that are within distance $r_a - 1$ from vertex $a$. Observe that
\[ G(a;r_a) \supset G_{V_D}(a;r_a) \supset G(a;r_a - 1) \]

\begin{lemma}
\label{le:iterconv} Let $D_1$ and $D_2$ be two different subsets
of $C$. Let $\{h^{j \to b}\}, \{g^{j \to b}\}$ be the unique fixed
points of the iterative equations on the graphs $G_{V_{D_1}}$ and
$G_{V_{D_2}}$. Then
\begin{equation*}
\begin{aligned}
& | h^{i \to a} - g^{i \to a} | \\
& \quad  \leq \left(\f{\Delta}{e^{-2\bt}+\Delta}\right)^{(r_a -
1)/2} \f{1}{\bt} \log \left[ 1 + (\Delta-1)e^{2\bt} \right]
\end{aligned}
\end{equation*}
\end{lemma}

\medskip
\begin{proof}
Since $G_{V_{D_1}}(a;r_a), G_{V_{D_2}}(a;r_a) \supset G(a;r_a -
1)$ we can set $\{h^{j \to b}(0)\} = \{g^{j \to b}(0)\}$ for the
messages in the sub-graph $G(a;r_a - 2)$.

To bound the difference in the messages at the boundary depending
on choice of $D$, observe the following: Let $v$ be any vertex at
distance $r_a - 1$ from $a$ and let $u$ be any neighbor of $v$
that is at distance $r_a - 2$. Let $0 \leq \delta_1,\delta_2 \leq
\Delta$ denote the degree of the vertex $v$ in the graphs
$G_{V_{D_1}}(a;r_a), G_{V_{D_2}}(a;r_a)$ respectively. Let $e$
denote the edge joining $v$ to $u$.  Then from Lemma
\ref{le:bound} we see that
\[ | h^{e \to u} - g^{e \to u} | \leq \f{1}{\bt}
\log \left[ 1 + (\Delta-1)e^{2\bt} \right]. \]

We can use Lemma \ref{lem:Lip} to determine the propagation of
this difference to the messages at $a$. It is not difficult to see
from repeated use of Lemma \ref{lem:Lip} that
\begin{equation*}
 | h^{i \to a} - g^{i \to a} |   \quad  \leq
\left(\f{\Delta}{e^{-2\bt}+\Delta}\right)^{(r_a - 1)/2} \f{1}{\bt}
\log \left[ 1 + (\Delta-1)e^{2\bt} \right]
\end{equation*}
\end{proof}
For simplicity of notation let us denote
\[ \nu = \f{1}{\bt} \log \left[ 1 + (\Delta-1)e^{2\bt} \right], \quad K = \f{\Delta}{e^{-2\bt}+\Delta}.   \]
It is easy to see that $\nu \leq 3$ for large enough $\beta$.

Let $i$ be an edge that is connected to the vertex $a$. We will
show that the free-energy shift $\Delta F_i$ for the graph $G$ can
be approximated by the free energy shift corresponding to the tree
$G(a,r_a)$.

\begin{lemma}
\label{le:lemfe} Let $\delta =  \nu K^{(r_a - 1)/2}$. Then,
\[  e^{-2\bt \delta } \frac{Z(G(a;r_a))}{Z(G_i(a;r_a))} \leq \frac{Z(G)}{Z(G_i)}
\leq e^{2\bt \delta} \frac{Z(G(a;r_a))}{Z(G_i(a;r_a))}
\]
\end{lemma}

\medskip

\begin{proof}
Observe that
\[ Z(G) = \sum_D Z(G_{V_D}(a;r_a))Z(H_{U_D}(a;r_a)) \]
Similarly for $G_i$, obtained by removing edge $i$ in $G$, we
obtain
\[ Z(G_i) = \sum_D Z(G_{V_D,i}(a;r_a))Z(H_{U_D}(a;r_a)) \]

Let $h_\emptyset^{i \to a}$ be the converged values of the
iterative equations for the case when $D = \emptyset$. Combining
the result for trees and our bounds on the converged values of
$h^{i \to a}$ for different initial conditions in Lemma
\ref{le:iterconv} we know that
\begin{equation*}
1 + e^{\beta(h_\emptyset^{i \to a} + h_\emptyset^{i \to b} -2
\delta)}  \leq \frac{Z(G_{V_D}(a;r_a))}{Z(G_{V_D,i}(a;r_a))} \leq
1 + e^{\beta(h_\emptyset^{i \to a} + h_\emptyset^{i \to b} +
2\delta )}.
\end{equation*}

Therefore for all choices of $D$ the following holds

\[\frac{Z(G(a;r_a))}{Z(G_i(a;r_a))}e^{-2\beta\delta}
\le \frac{Z(G_{V_D}(a;r_a))}{Z(G_{V_D,i}(a;r_a))}\leq
\frac{Z(G(a;r_a))}{Z(G_i(a;r_a))}e^{2\beta\delta}\]

Using this result and the decompositions of $Z(G)$ and $Z(G_i)$
presented above we obtain that
\[\frac{Z(G(a;r_a))}{Z(G_i(a;r_a))}e^{-2\beta\delta}
\leq \frac{Z(G)}{Z(G_i)}\leq
\frac{Z(G(a;r_a))}{Z(G_i(a;r_a))}e^{2\beta\delta}\]

\end{proof}

Observing that the boundary conditions do not influence the value
of $h^{i \to a}$, the fixed points of the iterative equations for
the graph $G$, we can infer that

\begin{equation*}
\left(1 + e^{\beta(h{i \to a} + h{i \to b})}\right)
e^{-2\beta\delta}  \leq \frac{Z(G)}{Z(G_i)}   \leq \left(1 +
e^{\beta(h{i \to a} + h{i \to b})}\right) e^{2\beta\delta}.
\end{equation*}
In a very similar manner to the removal of an edge, we can also
show that for some $\tilde{\delta} \leq \log(\Delta) + \delta$
\begin{equation}
\label{eq:varremo}
\frac{Z(G(a;r_a))}{Z(G_a(a;r_a))}e^{-\beta\tilde{\delta}} \leq
\frac{Z(G)}{Z(G_a)}\leq
\frac{Z(G(a;r_a))}{Z(G_a(a;r_a))}e^{\beta\tilde{\delta}}.
\end{equation}
Similar to the case of the removal of the edge, the above equation
implies that
\begin{equation*}
\left( e^{-\bt} + \sum_{i \in E(a)} e^{\bt h^{i \to a}} \right)
e^{-\beta\tilde{\delta}}  \leq \frac{Z(G)}{Z(G_a)}  \leq
\left(e^{-\bt} + \sum_{i \in E(a)} e^{\bt h^{i \to a}} \right)
e^{\beta\tilde{\delta}}.
\end{equation*}

\medskip

\subsection{Proof of Claim \ref{cl:cavity} for graphs with large
girth} ~

\medskip

Let $G$ be a graph that satisfies the {\em girth condition}.
\begin{equation}
\textrm{girth}(G)>\f{8(\log\nu+\log\bt+2\log m+ \log \Delta +
\log\f{1}{\log(1+\ep)})}{\log\f{1}{K}} \label{eq:girthcond}
\end{equation}
where $m = |V| + |E|$.

\begin{theorem}\label{th:mainform}
For any $\ep,\bt>0$. Let $G$ be a graph satisfying the girth
condition. Then the free energy shifts approximates $Z_G$ within
factor $1+\ep$, i.e.
\begin{equation}
\label{eq:feshftForm} Z_G(1+\ep)^{-1}\leq X_G \triangleq\f{\prod_a
\f{Z_G}{Z_{G_a}}}{\prod_i\f{Z_G}{Z_{G_i}}} \leq Z_G(1+\ep).
\end{equation}

\end{theorem}

\medskip

Note that when $E=\emptyset$, then $X_G=1$ and so
\eqref{eq:feshftForm} holds. Let $E=\{i_1,i_2,\ldots,i_{|E|}\}$ be
an ordering of edges and $g= \lfloor \textrm{girth}(G)/4 - 1
\rfloor$. The following lemma is crucial to prove Theorem
\ref{th:mainform}.

\begin{lemma}\label{le:approxZbyX}
For all $1\leq r\leq |E|$:
\begin{equation*}
\frac{Z_{G_{E_r}}}{Z_{G_{E_{r-1}}}}e^{-\nu \Delta\bt mK^{g/2}}\leq
\frac{X_{G_{E_r}}}{X_{G_{E_{r-1}}}}\leq
\frac{Z_{G_{E_r}}}{Z_{G_{E_{r-1}}}}e^{\nu \Delta\bt mK^{g/2}}
\end{equation*}
where $E_r=\{i_1,\ldots,i_r\}$, $K=\f{\Delta}{\Delta+e^{-2\bt}}$
and $m=n+|E|$.
\end{lemma}

Before proving Lemma \ref{le:approxZbyX} we will show how it can
be used to prove Theorem \ref{th:mainform}.

\medskip

\begin{proof} [Proof of Theorem \ref{th:mainform}]

Assuming Lemma \ref{le:approxZbyX} and taking the telescopic
product of $r = 1$ to $|E|$, we obtain
\begin{equation*}
e^{-\nu \Delta\bt
m^2K^{g/2}}\prod_{r=1}^{|E|}\frac{Z_{G_{E_r}}}{Z_{G_{E_{r-1}}}}
\leq\prod_{r=1}^{|E|}\frac{X_{G_{E_r}}}{X_{G_{E_{r-1}}}} \leq
e^{\nu \Delta\bt
m^2K^{g/2}}\prod_{r=1}^{|E|}\frac{Z_{G_{E_r}}}{Z_{G_{E_{r-1}}}}
\end{equation*}
and hence
\[
e^{-\nu \Delta\bt
m^2\left(\f{\Delta}{\Delta+e^{-2\bt}}\right)^{g/2}}Z_G\leq X_G
\leq e^{\nu \Delta\bt
m^2\left(\f{\Delta}{\Delta+e^{-2\bt}}\right)^{g/2}}Z_G.
\]
The assumptions on the girth of the graph in \eqref{eq:girthcond}
implies that  $e^{\nu \Delta\bt m^2K^{g/2}}\leq 1+\ep$ and this
completes the proof.
\end{proof}

\medskip

Therefore, all that remains is to prove Lemma \ref{le:approxZbyX}.

\medskip

\begin{proof}[Proof of Lemma \ref{le:approxZbyX}]

We need to show that
\begin{equation*}
\frac{Z_{G_{E_r}}}{Z_{G_{E_{r-1}}}}e^{-\nu \Delta\bt mK^{g/2}}\leq
\frac{X_{G_{E_r}}}{X_{G_{E_{r-1}}}}\leq
\frac{Z_{G_{E_r}}}{Z_{G_{E_{r-1}}}}e^{\nu \Delta\bt mK^{g/2}}.
\end{equation*}

Observe that,
\begin{equation}
\frac{X_{G_{E_r}}}{X_{G_{E_{r-1}}}} = \f{\prod_{a\in V}
\f{Z_{G_{E_r}}}{Z_{G_{E_r,a}}}}{\prod_{i\in E\stm
E_r}\f{Z_{G_{E_r}}}{Z_{G_{E_r,i}}}}\left(\f{\prod_{a\in V}
\f{Z_{G_{E_{r-1}}}}{Z_{G_{E_{r-1},a}}}}{\prod_{i\in E\stm
E_{r-1}}\f{Z_{G_{E_{r-1}}}}{Z_{G_{E_{r-1},i}}}}\right)^{-1}
=\f{\prod_{a\in V}
\f{Z_{G_{E_r}}}{Z_{G_{E_{r-1}}}}\left(\f{Z_{G_{E_r,a}}}{Z_{G_{E_{r-1},a}}}\right)^{-1}}{\prod_{i\in
E\stm
E_r}\f{Z_{G_{E_r}}}{Z_{G_{E_{r-1}}}}\left(\f{Z_{G_{E_r,i}}}{Z_{G_{E_{r-1},i}}}\right)^{-1}}.
\label{eq:feshftFormExp}
\end{equation}

Let $a_r$ be an endpoint of the edge $i_r$. We will estimate the
product
\[ \f{\prod_{a\in V}
\f{Z_{G_{E_r}}}{Z_{G_{E_{r-1}}}}\left(\f{Z_{G_{E_r,a}}}{Z_{G_{E_{r-1},a}}}\right)^{-1}}{\prod_{i\in
E\stm
E_r}\f{Z_{G_{E_r}}}{Z_{G_{E_{r-1}}}}\left(\f{Z_{G_{E_r,i}}}{Z_{G_{E_{r-1},i}}}\right)^{-1}}
\]
by partitioning the vertices and edges into two groups; those that
are in $G(a_r,g)$ and those that are outside $G(a_r,g)$.

Using equation \eqref{eq:varremo} whenever $a\notin
G_{E_{r-1}}(a_r;g)$, we have
\begin{equation}
e^{-\nu \Delta\bt
K^{g/2}}\leq\f{Z_{G_{E_r}}}{Z_{G_{E_{r-1}}}}\left(\f{Z_{G_{E_r,a}}}{Z_{G_{E_{r-1},a}}}\right)^{-1}\leq
e^{\nu \Delta\bt K^{g/2}}.\label{eq:VrtShftApp}
\end{equation}

Similarly whenever $i\notin G_{E_{r-1}}(a_r;g)$ then from Lemma
\ref{le:lemfe} we have
\begin{equation}
e^{-\nu \Delta\bt
K^{g/2}}\leq\f{Z_{G_{E_r}}}{Z_{G_{E_{r-1}}}}\left(\f{Z_{G_{E_r,i}}}{Z_{G_{E_{r-1},i}}}\right)^{-1}\leq
e^{\nu \Delta\bt K^{g/2}}.\label{eq:EdgShftApp}
\end{equation}

Now we consider the case when $a,i \in G(a_r,g)$.

Since $G_{E_{r-1}}(a_r;g)\subset G_{E_{r-1}}(a_r;2g)$ and both
graphs are trees, we can use  \eqref{eq:varremo}  for $a\in
G_{E_{r-1}}(a_r;g)$ to obtain
\begin{equation}
\begin{aligned}
e^{-\nu \Delta\bt
K^{g/2}}\f{Z_{G_{E_r}}}{Z_{G_{E_r;a}}}&\leq\f{Z_{G_{E_r}(a_r;g)}}{Z_{G_{E_r,a}(a_r;g)}}\leq\f{Z_{G_{E_r}}}{Z_{G_{E_r,a}}}e^{\nu
\Delta\bt K^{g/2}}\\ e^{-\nu \Delta\bt
K^{g/2}}\f{Z_{G_{E_{r-1}}}}{Z_{G_{E_{r-1},a}}}&\leq\f{Z_{G_{E_{r-1}}(a_r;g)}}{Z_{G_{E_{r-1},a}(a_r;g)}}\leq\f{Z_{G_{E_{r-1}}}}{Z_{G_{E_{r-1},a}}}e^{\nu
\Delta\bt K^{g/2}}.
\end{aligned}
\label{eq:VrtShftAppTree}
\end{equation}

Similarly, for $i\in G_{E_{r-1}}(a_r;g)$ using Lemma
\ref{le:lemfe} we have
\begin{equation}
\begin{aligned}
e^{-\nu \Delta\bt
K^{g/2}}\f{Z_{G_{E_r}}}{Z_{G_{E_r,i}}}&\leq\f{Z_{G_{E_r}(a_r,g)}}{Z_{G_{E_r,i}(a_r,g)}}\leq\f{Z_{G_{E_r}}}{Z_{G_{E_r,i}}}e^{\nu
\Delta\bt
K^{g/2}}\\
e^{-\nu \Delta\bt
K^{g/2}}\f{Z_{G_{E_{r-1}}}}{Z_{G_{E_{r-1},i}}}&\leq\f{Z_{G_{E_{r-1}}(a_r,g)}}{Z_{G_{E_{r-1},i}(a_r,g)}}\leq\f{Z_{G_{E_{r-1}}}}{Z_{G_{E_{r-1},i}}}e^{\nu
\Delta\bt K^{g/2}}
\end{aligned}
\label{eq:EdgShftAppTree}
\end{equation}

Note that,

\begin{equation}
\frac{X_{G_{E_r}}}{X_{G_{E_{r-1}}}} = \f{\prod_{a\notin G(a_r;g)}
\f{Z_{G_{E_r}}}{Z_{G_{E_{r-1}}}}\left(\f{Z_{G_{E_r,a}}}{Z_{G_{E_{r-1},a}}}\right)^{-1}}{\prod_{i
\notin
G(a_r;g)}\f{Z_{G_{E_r}}}{Z_{G_{E_{r-1}}}}\left(\f{Z_{G_{E_r,i}}}{Z_{G_{E_{r-1},i}}}\right)^{-1}}
\f{\prod_{a\in G(a_r;g)}
\f{Z_{G_{E_r}}}{Z_{G_{E_r,a}}}}{\prod_{i\in
G(a_r;g)}\f{Z_{G_{E_r}}}{Z_{G_{E_r,i}}}}\left(\f{\prod_{a\in
G(a_r;g)} \f{Z_{G_{E_{r-1}}}}{Z_{G_{E_{r-1},a}}}}{\prod_{i\in
G(a_r;g)}\f{Z_{G_{E_{r-1}}}}{Z_{G_{E_{r-1},i}}}}\right)^{-1} .
\label{eq:feshftFormExp1}
\end{equation}

Using equations \eqref{eq:VrtShftApp}, \eqref{eq:EdgShftApp}, in
the first product and equations \eqref{eq:VrtShftAppTree},
\eqref{eq:EdgShftAppTree} in the second product (and from the
definition of $X_G$), we obtain
\begin{equation*}
\begin{aligned}
e^{-\nu \Delta\bt
mK^{g/2}}\frac{X_{G_{E_r}(a_r;g)}}{X_{G_{E_{r-1}}(a_r;g)}}\leq
\frac{X_{G_{E_r}}}{X_{G_{E_{r-1}}}}\leq e^{\nu \Delta\bt
mK^{g/2}}\frac{X_{G_{E_r}(a_r;g)}}{X_{G_{E_{r-1}}(a_r;g)}}.
\end{aligned}
\end{equation*}

Lemma \ref{le:fshfttree} implies that $X_G = Z_G$ when $G$ is a
tree. Therefore, since $G_{E_r}(a_r;g)$, $G_{E_{r-1}}(a_r;g)$ are
trees we can replace
$\frac{X_{G_{E_r}(a_r;g)}}{X_{G_{E_{r-1}}(a_r;g)}}$ with
$\frac{Z_{G_{E_r}(a_r;g)}}{Z_{G_{E_{r-1}}(a_r;g)}}$ and
subsequently using Lemma \ref{le:lemfe} to replace
$\frac{Z_{G_{E_r}(a_r;g)}}{Z_{G_{E_{r-1}}(a_r;g)}}$ with
$\frac{Z_{G_{E_r}}}{Z_{G_{E_{r-1}}}}$, we complete the proof of
Lemma \ref{le:approxZbyX}.
\end{proof}

\section{CONCLUSIONS AND FUTURE WORKS}

In this paper we show the validity of the cavity method for the
problem of counting the number of matchings for graphs with large
girth. The girth condition we have in this paper is quite
restrictive and several graphs of practical relevance do not meet
this condition. However we hope that the methods presented here
can be extended in a straightforward manner to random regular
graphs and Erd\"{o}s-R\'{e}nyi graphs. This would lead to, as
observed in \cite{ZdeborovaMezard}, tighter estimates for counting
the number of matchings in such graphs.

We also demonstrate the convergence and uniqueness of the
iterative equations for arbitrary graphs. The techniques used in
this paper do not heavily depend on the nature of the problem and
therefore there is a good possibility of these techniques having a
wider interest and applicability to other important problems were
cavity method has been applied.






\end{document}